\documentclass[11pt]{article}

\usepackage{amsmath, amssymb}
\usepackage{amsthm}

\newtheorem{theorem}{Theorem}
\newtheorem{lemma}[theorem]{Lemma}
\newtheorem{corollary}[theorem]{Corollary}
\theoremstyle{definition}
\newtheorem{remark}[theorem]{Remark}

\newcommand{\N}{\mathbb{N}}
\newcommand{\B}{\mathbb{B}}
\newcommand{\Per}{\mathcal{P}}

\newcommand{\AP}{\mathcal{AP}}
\newcommand{\SAP}{\mathcal{SAP}}
\newcommand{\EAP}{\mathcal{EAP}}
\newcommand{\p}{\mathop{\mathrm{pr}}}
\renewcommand{\r}{\mathop{\mathrm R}} 
\newcommand{\s}{\mathop{\mathrm s}}   
\renewcommand{\b}{\mathop{\mathrm b}} 

\title{
Almost Periodicity, Finite Automata Mappings\\
and Related Effectiveness Issues}
\author{Yuri Pritykin%
\thanks{Moscow State University, Russia, e-mail: \texttt{pritykin@lpcs.math.msu.su},
\texttt{yura@mccme.ru}. The work was partially supported by RFBR
grants 06-01-00122, 05-01-02803, Scientific Schools grant 358.2003.1
and Kolmogorov grant of Institute of New Technologies.}}

\date{\today}

\begin{document}

\maketitle

\begin{abstract}
The paper studies different variants of almost periodicity notion.
We introduce the class of eventually strongly almost periodic
sequences where some suffix is strongly almost periodic (=uniformly
recurrent). The class of almost periodic sequences includes the
class of eventually strongly almost periodic sequences, and we prove
this inclusion to be strict. We prove that the class of eventually
strongly almost periodic sequences is closed under finite automata
mappings and finite transducers. Moreover, an effective form of this
result is presented. Finally we consider some algorithmic questions
concerning almost periodicity.
\end{abstract}

\section{Introduction}
\label{Introduction}

Strongly almost periodic sequences (=uniformly recurrent infinite
words) were studied in the works of Morse and Hedlund
\cite{Symb01,Symb02} and of many others (e.~g., see
\cite{Cass,MuchSemUsh}). A~sequence is strongly almost periodic if
every its factor occurs infinitely many times with bounded
distances. This notion first appeared in the field of symbolic
dynamics, but then turned out to be interesting in connection with
computer science, mathematical logic, combinatorics on words. Almost
periodic sequences were introduced in~\cite{SemArithm} while
studying logical theories of unary functions over~$\N$. A~sequence
is almost periodic if every its factor either occurs infinitely many
times with bounded distances or occurs only finitely many times. We
introduce a new class of sequences called eventually strongly almost
periodic, where some suffix is strongly almost periodic. Then we
study some properties of this class.

This paper is organized as follows.

In Section~\ref{Almost_periodicity} we give the formal definitions
of different generalizations of periodicity notion. The class of
almost periodic sequences includes the class of eventually strongly
almost periodic sequences. We prove this inclusion to be strict
(Theorem \ref{alPer-NotEssAlPer}).

Section~\ref{Finite_automata_mappings} concerns automata mappings.
Almost periodic sequences were studied in detail
in~\cite{SemLog,MuchSemUsh}. In particular, the authors prove that
the class of almost periodic sequences is closed under finite
automata mappings (=mappings done by synchronizing finite
transducers). Evidently, the class of finite automata mappings of
strongly almost periodic sequences contains the class of eventually
strongly almost periodic sequences. The main result of the article
(Theorem~\ref{finiteAutomataStrong}) states the equality of the
classes. In other words, Theorem~\ref{finiteAutomataStrong} says
that finite automata preserve the property of eventual strong almost
periodicity. Moreover, an effective variant of this theorem is
proved (Theorem~\ref{finiteAutomataStrong-effective}). Then we
consider a generalized version of finite automaton, i.~e., finite
transducer, and prove the same statement for it.

In Section~\ref{Effectiveness} we deal with some algorithmic
questions connected with almost periodicity. Namely, we prove that
some problems or properties do not have corresponding effective
analogs (in contrast to Theorem~\ref{finiteAutomataStrong} with
effective version in Theorem~\ref{finiteAutomataStrong-effective}).
For instance, we prove that given eventually strongly almost
periodic sequence and its regulator we can not find any prefix which
is enough to cut to obtain strongly almost periodic sequence.

Let us introduce some basic notions and notations.

Denote $\{0,1\}$ by $\B$, the set of nonnegative integers $\{0, 1,
2,\dots\}$ by $\N$. Let $\Sigma$ be a finite alphabet with at least
two symbols. We consider the sequences over this alphabet, i.~e.,
the mappings $\omega\colon\N\to \Sigma$. The set of all such
sequences forms Cantor metric space. Denote this space
by~$\Sigma^\N$. Then $\lim_{n\to\infty}x_n = \omega$, if $\forall i\
\exists n\ \forall m > n\ x_m(i) = \omega(i)$ (this definition works
for finite~$x_n$ too).

Denote by $\Sigma^*$ the set of all finite strings over $\Sigma$
including the empty string~$\Lambda$. If $i\le j$ are nonnegative
integers, denote by $[i,j]$ the segment of $\N$ with ends in~$i$
and~$j$, i.~e., the set $\{i, i+1, i+2,\dots,j\}$. Also denote by
$\omega[i,j]$ a substring $\omega(i)\omega(i+1)\dots\omega(j)$ of a
sequence $\omega$. A segment $[i,j]$ is an occurrence of a string
$x\in \Sigma^*$ in a sequence $\omega$ if $\omega[i,j]=x$. We say
that $x\ne\Lambda$ is a factor of $\omega$ if $x$ occurs in
$\omega$. The string of the form $\omega[0,i]$ for some $i$ is
called a prefix of $\omega$, and respectively the sequence of the
form $\omega(i)\omega(i+1)\omega(i+2)\dots$ for some $i$ is called a
suffix of $\omega$ and is denoted by $\omega[i,\infty)$. Denote by
$|x|$ the length of a string~$x$. The occurrence $x = \omega[i,j]$
in $\omega$ is $k$-aligned if $k|i$. Imagining the sequences going
horizontally from the left to the right, we use terms ``to the
right'' and ``to the left'' to talk about greater and smaller
indices respectively.

\section{Almost periodicity}
\label{Almost_periodicity}

A sequence $\omega$ is periodic if for some $T$ we have $\omega(i) =
\omega(i + T)$ for each $i\in\N$. This $T$ is called a period of
$\omega$. The class of all periodic sequences we denote by~$\Per$.
Let us consider some extensions of this class.

A sequence $\omega$ is called \emph{almost periodic} if for any
factor~$x$ of $\omega$ occurring in it infinitely many times there
exists a number~$l$ such that any factor of $\omega$ of length~$l$
contains at least one occurrence of~$x$. We denote the class of all
almost periodic sequences by $\AP$.

A sequence $\omega$ is called \emph{strongly almost periodic} if for
any factor~$x$ of $\omega$ there exists a number~$l$ such that any
factor of $\omega$ of length~$l$ contains at least one occurrence
of~$x$ (and therefore $x$ occurs in $\omega$ infinitely many times).
Obviously, to show strong almost periodicity of a sequence it is
sufficient to check the mentioned condition only for all prefixes
but not for all factors. Denote by $\SAP$ the class of all strongly
almost periodic sequences.

We also introduce an additional definition: a sequence $\omega$ is
\emph{eventually strongly almost periodic} if some its suffix is
strongly almost periodic. The class of all eventually strongly
almost periodic sequences we denote by $\EAP$.

Suppose $\omega\in\EAP$. Denote by~$\p(\omega)$ the minimal $n$ such
that $\omega[n,\infty)\in\SAP$. Thus for each $m \geqslant
\p(\omega)$ we have $\omega[m,\infty)\in\SAP$.

A function $\r_\omega\colon\N\to\N$ is an \emph{almost periodicity
regulator} of a sequence $\omega\in\AP$, if\\[1mm]
(1) every string of length $n$ occurring in $\omega$ infinitely many
times, occurs on any factor of length $\r_\omega(n)$ in $\omega$;\\
(2) any string of length $n$ occurring finitely many times in
$\omega$, does not occur in $\omega[\r_\omega(n),\infty)$.\\[1mm]
The latter condition is important only for sequences in
$\AP\setminus\SAP$. Notice that regulator is not unique: any
function greater than regulator is also a regulator.

Obviously, $\Per\subset\SAP\subset\EAP\subset\AP$. In fact, all
these inclusions are strict. For instance the famous Thue--Morse
sequence $\omega_T = 0110100110010110\dots$
(see~\cite{Thue,ThueMorse} or Section~\ref{Effectiveness}) is an
example of the element in $\SAP$ but not in $\Per$ (moreover, $\SAP$
has cardinality continuum while $\Per$ is countable, see~\cite{Jac}
or~\cite{MuchSemUsh} for proofs). The inequality $\SAP \subsetneq
\EAP$ is obvious. Let us prove $\EAP \subsetneq \AP$.

\begin{theorem}
\label{alPer-NotEssAlPer} There exists a binary sequence
$\omega\in\AP\setminus\EAP$.
\end{theorem}
\begin{proof}
Construct a sequence of binary strings $a_0=1$, $a_1=10011$,\\
$a_2=1001101100011001001110011$, and so on, by this rule:
  $$
 a_{n+1}=a_n\overline a_n\overline a_na_na_n,
  $$
where $\overline x$ is a string obtained from $x$ by changing every
0 to 1 and vice versa. Put
  $$
 c_n = a_na_na_na_n
  $$
and
  $$
 \omega = c_0c_1c_2c_3\dots
  $$
Let us prove that $\omega\in\AP\setminus\EAP$.

The length of $a_n$ is $5^n$, so the length of $c_0c_1\dots c_{n-1}$
is $4(1 + 5 + \dots + 5^{n-1}) = 5^n-1$. By definition, put
  $$
 l_n = 5^n-1 = |c_0c_1\dots c_{n-1}|.
  $$

Let us show that $\omega$ is almost periodic. Suppose $x\ne\Lambda$
occurs in $\omega$ infinitely many times. Take $n$ such that
$|x|<5^n$. Suppose $[i,j]$ is an occurrence of $x$ in $\omega$ such
that $i\ge l_n$. By construction, for any $k$ we can consider
$\omega[l_k,\infty)$ as a concatenation of strings $a_k$ and
$\overline a_k$. Thus (by assumption about $i$) the string $x$ is a
substring of either $a_na_n$, $a_n\overline a_n$, $\overline a_na_n$
or $\overline a_n\overline a_n$. Notice that 10011 contains all
strings of length two (00, 01, 10 and 11), so $a_{n+1}$ contains
each of $a_na_n$, $a_n\overline a_n$, $\overline a_na_n$, $\overline
a_n\overline a_n$. Hence $x$ is a substring of $a_{n+1}$. Similarly,
$x$ is a substring of $\overline a_{n+1}$. In each factor of length
$2|a_{n+1}|$ of $\omega[l_{n+1},\infty)$, $a_{n+1}$ or $\overline
a_{n+1}$ occurs. Hence for $l = (5^{n + 1} - 1) + 2\cdot5^{n+1}$ the
string $x$ occurs on every factor of length~$l$ in~$\omega$.

Now let us prove that for any $n>0$ the string $c_n$ does not occur
in $\omega[l_{n+1},\infty)$. This implies that $c_n$ occurs finitely
many times in the suffix $\omega[l_n,\infty)$, i.~e., this suffix is
not strongly almost periodic. Therefore $\omega$ is not eventually
strongly almost periodic.

Let $\nu = \omega[l_{n+1},\infty)$. As above, for each $k$, $1\le
k\le n+1$, $\nu$ is a concatenation of strings $a_k$ and $\overline
a_k$. Assume $c_n$ occurs in $\nu$ and let $[i,j]$ be one of this
occurrences. For $n>0$ the string $c_n$ begins with $a_1$, hence
$[i,i+4]$ is an occurrence of $a_1$ in~$\nu$. We see that $a_1 =
10011$ occurs in $a_1a_1 = 1001110011$, $a_1\overline a_1 =
1001101100$, $\overline a_1a_1 = 0110010011$ or $\overline
a_1\overline a_1 = 0110001100$ only in 0th or 5th position. Thus
$[i, j]$ is 5-aligned, hence $\nu$ and $c_n$ can be considered as
constructed of ``letters'' $a_1$ and $\overline a_1$, and we assume
that $c_n$ occurs in $\nu$. Now it is easy to prove by induction on
$m$ that $[i,j]$ is $5^m$-aligned for $1\leqslant m\leqslant n$,
i.~e., we can consider $\nu$ and $c_n$ to be constructed of
``letters'' $a_m$ and $\overline a_m$, and assume that $c_n$ occurs
in $\nu$ (the base for $m = 1$ is already proved, and we can repeat
the same argument changing 1 and 0 to $a_m$ and $\overline a_m$ and
taking into account that $c_n$ begins with $a_m$ for each $1\le m\le
n$).

Therefore we have shown that $[i,j]$ is $5^n$-aligned, hence if we
consider $\nu$ and $c_n$ to be constructed of ``letters'' $a_n$ and
$\overline a_n$, then $c_n = a_na_na_na_n$ occurs in $\nu$. But
notice that in any sequence constructed by concatenation of strings
$a_1 = 10011$ and $\overline a_1 = 01100$ there is no any occurrence
of 0000 or 1111. That is why $c_n$ also can not occur in $\nu$. This
is a contradiction.
\end{proof}

Moreover, it is quite easy to modify the proof and to show that
$\AP\setminus\EAP$ has cardinality continuum. For instance for each
sequence $\tau\colon\N\to\{4,5\}$ we can construct $\omega_\tau$ in
the same way as in the proof of Theorem \ref{alPer-NotEssAlPer}, but
instead of $c_n$ we take
  $$
 c_n^{(\tau)}=\underbrace{a_na_n\dots a_n}_{\tau(n)}.
  $$
Obviously, all $\omega_\tau$ are different for different~$\tau$
and hence there exists continuum of various~$\tau$.

\section{Finite automata mappings}
\label{Finite_automata_mappings}

It seems interesting to understand whether some transformations of
sequences preserve the property of almost periodicity. The simplest
type of algorithmic transformation is finite automaton mapping.
Another motivation, less philosophical, is that finite automata
mappings were one of the most useful tools in \cite{SemLog} while
studying almost periodicity and finding some criterion for
first-order and monadic theories of unary functions over~$\N$ to be
decidable.

\emph{Finite automaton} is a tuple $F = \langle \Sigma, \Delta, Q,
\tilde q, f\rangle$, where $\Sigma$ and $\Delta$ are finite sets
called input and output alphabets respectively, $Q$ is a finite set
of states, $\tilde q\in Q$ is the initial state, and
  $$
 f\colon Q\times \Sigma\to Q\times \Delta
  $$
is the transition function. For $\alpha\in\Sigma^\N$ consider the
sequence $\langle p_n, \beta(n)\rangle_{n=0}^\infty$, where $p_n\in
Q$, $\beta(n)\in \Delta$, and assume $p_0=\tilde q$ and $\langle
p_{n+1}, \beta(n)\rangle = f(p_n, \alpha(n))$ for each~$n$. Then we
call $\beta = F(\alpha)$ a finite automata mapping of~$\alpha$. If
$[i,j]$ is an occurrence of a string $x$ in~$\alpha$, and $p_i=q$,
then we say that automaton $F$ comes to this occurrence of $x$ being
in the state~$q$.

In~\cite{SemLog,MuchSemUsh} the following statement was proved.

\begin{theorem}\label{finiteAutomataWeak}
If $F$ is a finite automaton and $\omega\in\AP$, then
$F(\omega)\in\AP$.
\end{theorem}

We can prove a counterpart of this statement for eventually strongly
almost periodic sequences.

\begin{theorem}\label{finiteAutomataStrong}
If $F$ is a finite automaton and $\omega\in\EAP$, then
$F(\omega)\in\EAP$.
\end{theorem}
\begin{proof}
Obviously, it is enough to prove the theorem for $\omega\in\SAP$,
since prefix does not matter.

Thus let $\omega\in\SAP$. By Theorem~\ref{finiteAutomataWeak},
$F(\omega)\in\AP$. Suppose $F(\omega)$ is not eventually strongly
almost periodic. It means that for any natural $N$ there exists a
string that occurs in $F(\omega)$ after position $N$ and does not
occur after that. Indeed, if we remove the prefix $[0,N]$ from
$F(\omega)$, we do not get strongly almost periodic sequence, hence
there exists a string occurring in this sequence only finitely many
times. Then take its rightmost occurrence.

Let $[i_0,j_0]$ be the rightmost occurrence of a string $y_0$ in
$F(\omega)$. For some $l_0$ the string $x_0 = \omega[i_0,j_0]$
occurs in every factor of the length $l_0$ in~$\omega$ (by the
property of strong almost periodicity). If $F$ comes to $i_0$ in the
state~$q_0$, then $F$ never comes to righter occurrences of $x$ in
the state $q_0$ because in this case automaton outputs $y_0$
completely.

Now let $[r,s]$ be the rightmost occurrence of some string $a$ in
$F(\omega)$, where $r>i_0+l_0$. On the factor $\omega[r-l_0,r]$
there exists an occurrence $[r',s']$ of the string $x_0$. By
definition of $r$ we have $r'>i_0$. Thus assume
  $$
 i_1=r',\ j_1=s,\ x_1=\omega[i_1,j_1],\ y_1=F(\omega)[i_1,j_1].
  $$
Since $a$ does not occur in $F(\omega)$ to the right of $r$, then
$y_1$ does not occur in $F(\omega)$ to the right of $i_1$, for it
contains $a$ as a substring. Therefore if the automaton comes to the
position $i_1$ in the state $q_1$, then it never comes to righter
occurrences of $x_1$ in the state $q_1$. Since $x_1$ begins with
$\omega[r',s']=x_0$, and $r'>i_0$, we get $q_1\ne q_0$. We have
found the string $x_1$ such that automaton $F$ never comes to
occurrences of $x_1$ to the right of $i_1$ in the state $q_0$ or
$q_1$.

Let $m = |Q|$. Arguing as above, for $k < m$ we construct the
strings $x_k = \omega[i_k,j_k]$ and corresponding different states
$q_k$, such that $F$ never comes to occurrences of $x_k$ in $\omega$
to the right of $i_k$ in the states $q_0,q_1,\dots,q_k$. For $k = m$
we have a contradiction.
\end{proof}

Notice that this proof is non-effective in the following sense.
Suppose we know $\omega\in\SAP$ and its almost periodicity
regulator~$\r_\omega$. Then by Theorem~\ref{finiteAutomataStrong}
some upper bound on $\p(F(\omega))$ exists for $F(\omega)\in\EAP$,
but the presented proof does not allow us to obtain any such bound.

Theorem~\ref{finiteAutomataStrong} was proved first
in~\cite{sapfintr}. The following effective version of this theorem
was announced in~\cite{conf06}.

For a function $g$ denote $\underbrace{g\circ g\circ \dots\circ
g}_n$ by $g^n$.

\begin{theorem}\label{finiteAutomataStrong-effective}
Let $F$ be a finite automaton with $n$ states and $\omega\in\SAP$.
Then $F(\omega)\in\EAP$ and
  $$
 \p(F(\omega)) \leqslant \r\nolimits_\omega^n(1) + \r\nolimits_\omega^{n-1}(1) +
 \dots + \r\nolimits_\omega(1).
  $$
\end{theorem}

To prove this theorem, first we consider particular type of automata
called reversible for which the statement of theorem is simple. Then
we introduce some construction in combinatorics on words which
allows us to reduce general situation to the case of reversible
automaton.

A finite automaton $F = \langle \Sigma, \Delta, Q, \tilde q,
f\rangle$ is \emph{reversible}, if for any $q\in Q$ and $a\in
\Sigma$ there exist unique $q'\in Q$ and $b\in \Delta$, such that
$f(q', a) = \langle q, b\rangle$. In other words, in such an
automaton each letter of the input alphabet $\Sigma$ performs a
permutation on $Q$ (output alphabet does not matter). If we have
some state, we can reconstruct the sequence of previous states from
the sequence of previous input letters (this is a reversibility
property).

\begin{theorem} \label{reversibleAutomata}
If $F$ is a reversible finite automaton and $\omega\in\SAP$, then
$F(\omega)\in\SAP$.
\end{theorem}
\begin{proof}
Suppose $x$ occurs in $\omega$, and $F$ comes to this occurrence in
the state $q$. Our goal is to prove that the next time when $F$
comes to $x$ in $\omega$ being in the state $q$, is at some distance
of the previous such situation, and we can also give upper bound for
this distance in terms of $|x|$ and $\r_\omega$. It means that the
same estimate for this distance works for any situation when $F$
comes to $x$ in the state $q$. So it is enough for our purpose.

Let $x_0 = \omega[r,s]$ be a factor of $\omega$; the automaton comes
to this occurrence in some state $q$. Let $[i_0,j_0]$ be the next
occurrence of $x_0$ in $\omega$, then $j_0 \leqslant r +
\r_\omega(|x|) + 1$. If $F$ comes to this occurrence being in the
state $q$, then all is done. Otherwise $F$ comes to $i_0$ being in
the state~$q_0 \ne q$. Let $x_1 = \omega[r,j_0]$, and suppose
$[i_1,j_1]$ is the next occurrence of $x_1$ in $\omega$, so $j_1
\leqslant r + \r_\omega(\r_\omega(|x|) + 1) + 1$. Suppose the
automaton comes to the position $i_1 + i_0$ being in the state
$q_1$. If $q_1 = q$, then all is done for $\omega[i_1 + i_0, j_1] =
x_0$. If $q_1 = q_0$, then $F$ comes to the position $i_1$ in the
state $q$, since $F$ is reversible, and all is done again. If it is
not the case, $q_1\ne q$ and $q_1\ne q_0$.

Similarly, for $x_2 = \omega[r,j_1]$ and its occurrence $[i_2, j_2]$
in $\omega$, such that $i_2 > r$ and $j_2 \leqslant r +
\r_\omega(\r_\omega(\r_\omega(|x|) + 1) + 1) + 1$, either we are
done or $F$ comes to the position $i_2 + i_1 + i_0$ in the state
$q_2$ where $q_2\ne q$, $q_2\ne q_0$ and $q_2\ne q_1$. Arguing in
the same way, for $k < m = |Q|$ we construct the strings $x_0,
x_1,\dots, x_k$ with occurrences $[i_0,j_0], [i_1,j_1],\dots,
[i_k,j_k]$ and different states $q_0,q_1,\dots,q_{k-1}$, such that
in worst case $F$ can not come to the position $i_k + i_{k-1}
+\dots+ i_0$ in states $q, q_0, \dots, q_{k-1}$. Thus for $k = m$ we
are done for sure, and the estimate for distance will be
$f(f(\dots(|x|)\dots))$, where $f = \r_\omega + 1$ and the number of
iterations is~$m$.
\end{proof}

For $\omega\in\Sigma^\N$, $\nu\in\Delta^\N$ define
$\omega\times\nu\in(\Sigma\times \Delta)^\N$ such that
$(\omega\times\nu)(i) = \langle\omega(i),\nu(i)\rangle$.

\begin{corollary} \label{SAPtimesPer}
If $\omega\in\SAP$ and $\nu\in\Per$, then $\omega\times\nu\in\SAP$.
\end{corollary}
\begin{proof}
Operation ``$\times$'' with periodic sequence can be simulated by
cyclic finite automaton which is obviously reversible.
\end{proof}

\begin{remark}\label{open-times}
Let us formulate some open questions connected with
Corollary~\ref{SAPtimesPer}. Let $\omega,\nu\in\SAP$. Then it is
interesting to know what we can say about $\omega\times\nu$. It is
not difficult to construct $\omega\times\nu\notin\AP$ or
$\omega\times\nu\in\EAP\setminus\SAP$. Can we construct
$\omega\times\nu\in\AP\setminus\EAP$? Does there exist any criterion
to determine whether $\omega\times\nu\in\AP$? An example of
$\omega,\nu\in\AP$ with $\omega\times\nu\notin\AP$ can be found
in~\cite{MuchSemUsh}.
\end{remark}

Now consider the following construction. Let $\omega\in\Sigma^\N$,
and suppose $a\in\Sigma$ occurs in $\omega$ infinitely many times.
Cut $\omega$ on the blocks like $xa$, where
$x\in(\Sigma\setminus\{a\})^*$, i.~e., on the blocks containing a
symbol $a$ on the end and not containing any other occurrences
of~$a$. To make this we need to cut after each occurrence of~$a$. If
$a$ occurs in $\omega$ at bounded distance, then the number of all
such blocks is finite (for example, if $\omega\in\AP$, then the
length of blocks is not more then $\r_\omega(1)$). Encode these
blocks by symbols of some finite alphabet, denote this alphabet by
$\b_{a,\omega}(\Sigma)$. Thus we obtained a new sequence in this
alphabet from $\omega$. Delete the first symbol of this sequence.
The result is called an \emph{$a$-split} of $\omega$ and is denoted
by $\s_a(\omega)$. For example, 0-split of the sequence
3200122403100110\dots is (0)(12240)(310)(0)(110)\dots

\begin{lemma}\label{split}
Let $\omega\in\SAP$, and suppose $a\in\Sigma$ occurs in $\omega$.
Then $\s_a(\omega)\in\SAP$.
\end{lemma}
\begin{proof}
Let $k$ be the maximal length of the $a$-split blocks. Consider a
prefix $x$ of $\s_a(\omega)$. The corresponding string $y$ in
$\omega$ is not longer than $k|x|$. Let $z = ay$, $|z| \leqslant
k|x| + 1$. The string $z$ occurs in $\omega$. Therefore $z$ occurs
on any factor of length $l = \r_\omega(k|x| + 1)$ in $\omega$. The
first and the last symbols of $z$ are $a$, so every such occurrence
is well-aligned relative to $a$-split of $\omega$. Hence for any
occurrence of $z$ in $\omega$ there is an occurrence of $x$ in
$\s_a(\omega)$. Therefore $x$ occurs on each factor of length
$\r_\omega(k|x| + 1)$ in~$\s_a(\omega)$.
\end{proof}

\begin{remark}
In connection with Lemma~\ref{split} an interesting question
appears, a particular case of the question in
Remark~\ref{open-times}. Instead of splitting $\omega$ on blocks
with some fixed symbol at the end, we can split $\omega$ arbitrarily
on blocks of various lengths. When the result is strongly almost
periodic or just almost periodic?
\end{remark}

Now we can prove the promised theorem.

\begin{proof}[Proof of Theorem~\ref{finiteAutomataStrong-effective}]
Let $F=\langle \Sigma, \Delta, Q, \tilde q, f\rangle$ and $|Q| = n$.
We construct an algorithm to compute some
  $$
 l \geqslant \p(F(\omega)),
  $$
and at the same time we prove
  $$
 l \leqslant \r\nolimits_\omega^n(1) + \r\nolimits_\omega^{n-1}(1) +
 \dots + \r\nolimits_\omega(1).
  $$

Assume that any automaton in the proof has the maximum possible
output alphabet ``input alphabet''$\times$``the set of states''
(general case can be obtained from this by projection). For example
for $F$ this is $\Sigma \times Q$. Correspondingly the transition
function $f$ writes the pair of a current state and an input symbol
to the output. Further we omit the second part of the transition
function value (i.~e., for instance, instead of $f(p, a) = \langle
q, b\rangle$ we write just $f(p, a) = q$ with $f(p, a) = \langle q,
\langle p, a\rangle\rangle$ in mind).

Let $\omega_0 = \omega$. Suppose every symbol of $\Sigma$ occurs in
$\omega_0$, elsewise we restrict $F$ only on the symbols occurring
in $\omega_0$; to determine these symbols effectively we should read
first $\r_{\omega_0}(1)$ symbols of $\omega_0$.

If $F$ is reversible, by Theorem~\ref{reversibleAutomata} we get
$\p(F(\omega_0)) = 0$. Otherwise some symbol $a_0\in\Sigma$
accomplishes not one-to-one mapping of $Q$, so the set
  $$
 Q_1 = \{q\ :\ \exists q'\ \ f(q', a_0) = q\}
  $$
is the proper subset of~$Q$. Consider
  $$
 \omega_1 = \s\nolimits_{a_0}(\omega_0),
  $$
which is strongly almost periodic by Lemma~\ref{split}. Notice that
starting in any state on $\omega_0$, the automaton $F$ comes to any
block of $a_0$-split of $\omega_0$ in the state of the set~$Q_1$,
because every such block has~$a_0$ at the end.

Let us construct a new automaton $F_1$ (effectively by $F$). Let the
input alphabet of $F_1$ be $\b_{a_0,\omega_0}(\Sigma)$, the set of
states be $Q_1$, and the value of the transition function on
$x\in\b_{a_0,\omega_0}(\Sigma)$, $q\in Q_1$ be the output of $F$
starting in the state $q$ on the string $x$ written in symbols of
$\Sigma$. Let the initial state of $F_1$ be the state of $F$ after
the work on the prefix of $\omega$ until the first occurrence of
$a_0$ (the prefix which we delete obtaining $\s_{a_0}(\omega_0)$
from $\omega_0$). Now the work of $F_1$ on~$\omega_1$ simulates the
work of $F$ on $\omega_0$. Notice that $\omega_1$ is obtained from
$\omega$ by deleting not more than $\r_{\omega_0}(1)$ first symbols,
counting in the alphabet~$\Sigma$.

We have the sequence $\omega_1$ (in the alphabet more than initial)
and the automaton $F_1$ with the set of states less than initial. If
$F_1$ is not reversible, then we can repeat the procedure of the
last paragraph. Thus we obtain the sequence $\omega_2$ in some
alphabet $\b_{a_1,\omega_1}(\b_{a_0,\omega_0}(\Sigma))$, and the
automaton $F_2$ with the set of states less than previous, working
on $\omega_2$. The sequence $\omega_2$ is obtained from $\omega_1$
by deleting not more than $\r_{\omega_1}(1)$ first symbols, counting
in the alphabet $\b_{a_0,\omega_0}(\Sigma)$. Therefore $\omega_2$,
written in the initial alphabet $\Sigma$, is obtained from $\omega$
by deleting not more than $\r_{\omega_0}(\r_{\omega_0}(1)) +
\r_{\omega_0}(1)$ first symbols, counting in the alphabet~$\Sigma$.

An automaton with a single state (and with arbitrary input alphabet)
is always reversible. Hence after $k$ repetitions of described
procedure for some $k \leqslant n$, we get the situation when the
reversible automaton $F_k$ works on strongly almost periodic
sequence $\omega_k$ in some alphabet (after each repetition of the
procedure the number of states decreases). The symbols of this
alphabet code the blocks of initial sequence. Thus
$F_k(\omega_k)\in\SAP$.

Writing $\omega_k$ in the alphabet $\Sigma$, we get some suffix
$\omega'$ obtained from $\omega$ by deleting a prefix not longer
than
  $$
 \r\nolimits_\omega^k(1) + \r\nolimits_\omega^{k-1}(1) + \dots +
 \r\nolimits_\omega(1) \leqslant \r\nolimits_\omega^n(1) +
 \r\nolimits_\omega^{n-1}(1) + \dots + \r\nolimits_\omega(1).
  $$

It only remains to check why $F_k(\omega_k)\in\SAP$ implies
$F(\omega')\in\SAP$. Let us explain this in one simple case when the
automaton $F_1$ obtained after the first iteration of procedure is
reversible (the general situation can be reduced to this case by
induction). Then $\omega'$ is obtained from $\omega$ by deleting
first symbols until the first occurrence of $a_0$. Let the initial
state of $F_1$ (in which $F$ comes to $\omega'$) be $q$. To show
$F(\omega')\in\SAP$ it is necessary and sufficient to check whether
for any prefix of $\omega'$ the occurrences of its copies in
$\omega'$, in which $F$ comes in the state $q$, are quite regular,
i.~e. these copies occur on each factor of length $l$ for some $l$
(sufficient condition is obvious, necessary condition follows from
our requirement for automata always to output the pair
$\langle$input symbol, current state$\rangle$).

Let $x$ be a prefix of $\omega'$ ending by $a_0$ (arbitrary prefix
is contained in some such prefix). We can correctly split it on
blocks ending by $a_0$, denote this split by $y$. The automaton
$F_1$ is reversible, so $F_1(\omega_1)\in\SAP$. By the necessary and
sufficient condition of the previous paragraph $F_1$ comes to $y$ in
the state $q$ on each factor of length $t$ for some $t$. Every such
situation corresponds in $\omega'$ to coming $F$ to some occurrence
of $x$ in the state $q$, and this happens on each factor of
length~$tk$, where $k \leqslant \r_\omega(1)$ is the maximal length
of blocks.
\end{proof}

The upper bound on $\p(F(\omega))$ in
Theorem~\ref{finiteAutomataStrong-effective} can not be
significantly improved, as it follows from the construction
in~\cite{Raskin} after small modifications.

It is interesting that now we have two different proofs of
Theorem~\ref{finiteAutomataStrong}, and it seems that there is no
any connection between them.

The results about finite automata mappings can be extended on more
general class of mappings done by finite transducers.

Let $\Sigma$ and $\Delta$ be finite alphabets. The mapping $h\colon
\Sigma^*\to \Delta^*$ is called a \emph{homomorphism}, if for any
$u,v\in \Sigma^*$ we have $h(uv)=h(u)h(v)$. Clearly, any
homomorphism is fully determined by its values on single-letter
strings. Let $\omega\in\Sigma^\N$. By definition, put
  $$
 h(\omega)=h(\omega(0))h(\omega(1))h(\omega(2))\dots
  $$

Suppose $h\colon \Sigma^*\to \Delta^*$ is a homomorphism,
$\omega\in\Sigma^\N$ is almost periodic. In \cite{MuchSemUsh} it was
shown that if $h(\omega)$ is infinite, then it is almost periodic.
Thus obviously if $\omega$ is strongly almost periodic, and
$h(\omega)$ is infinite, then $h(\omega)$ is also strongly almost
periodic. Indeed, it is enough to show that any $v$ occurring in
$h(\omega)$ occurs infinitely many times. However there exists some
factor $u$ of $\omega$ such that $h(u)$ contains $v$, but by the
definition of strong almost periodicity $u$ occurs in $\omega$
infinitely many times. Evidently, for $\omega\in\EAP$ we have
$h(\omega)\in\EAP$, if $h(\omega)$ is infinite.

Now we modify the definition of finite automaton, allowing it to
output any string (including the empty one) over output alphabet
reading only one character from input. This modification is called
\emph{finite transducer} (see~\cite{FinTrans}). Formally, we only
change the definition of translation function. Now it has the form
  $$
 f\colon Q\times \Sigma\to Q\times \Delta^*.
  $$
If the sequence $\langle p_n, v_n\rangle_{n=0}^\infty$, where
$p_n\in Q$, $v_n\in \Delta^*$, is the mapping of $\alpha$, then the
output is the sequence $v_0v_1v_2\dots$

Actually, we can decompose the mapping done by finite transducer
into two: the first one is a finite automaton mapping and another is
a homomorphism. Each of these mappings preserves the class $\AP$, so
we get the corollary: finite transducers map almost periodic
sequences to almost periodic. Similarly, by Theorem
\ref{finiteAutomataStrong} and arguments above we also get the
following

\begin{corollary}
Let $F$ be a finite transducer, $\omega\in\EAP$. Suppose $F(\omega)$
is infinite. Then $F(\omega)\in\EAP$.
\end{corollary}

\section{Effectiveness}
\label{Effectiveness}

Lots of interesting algorithmic questions naturally appear in
connection with almost periodicity, i.~e., if one can check some
property or find some characteristic algorithmically being given a
sequence. Sometimes these questions are just effective issues for
corresponding noneffective results, for example
Theorem~\ref{finiteAutomataStrong-effective} is an effective variant
of Theorem~\ref{finiteAutomataStrong}. Further, we mainly deal with
the case when the answers on these questions are negative. We prove
that some properties do not have effective analogs.

Formally, we consider an algorithm with an oracle for a sequence on
input. This algorithm halts on every oracle and outputs a finite
binary string or any other constructive object. The main property of
such an algorithm is continuity: it outputs the answer on having
read only finite number of symbols from the sequence. Thus to prove
non-effectiveness we only need to show discontinuity. In fact, such
proofs are just concrete complicated combinatorial constructions
showing this discontinuity.

If we have only a sequence, then we can not recognize almost any
property about this sequence. For example it is even impossible to
understand whether the symbol 1 occurs in given binary sequence: if
an algorithm checks some finite number of symbols and all these
symbols are 0, then it can not guarantee that 1 does not occur
further. The question about algorithmic decidability becomes more
interesting if we allow to give on input some additional
information. In the case of almost periodic sequences it may be an
almost periodicity regulator.

It is easy to decode unambiguously functions $\N\to\N$ and also
pairs $\langle$sequence, function$\rangle$ by binary sequences. That
is why we can correctly consider algorithms with an almost periodic
sequence $\omega$ and its regulator~$f$ on input.

From this point of view the above problem can be solved effectively:
reading first $f(1)$ symbols of the sequence we can say whether or
not 1 occurs in it, and moreover reading next $f(1)$ symbols we can
say whether 1 occurs in it finitely or infinitely many times.

The following several theorems are examples of problems concerning
almost periodicity which do not have effective analogs. It is
especially interesting that the Theorem~\ref{EAPprefixNumber}
results are absolutely contrary to the results of
Theorem~\ref{finiteAutomataStrong-effective}.
Theorem~\ref{find_pr_by_EAP} is also connected with
Theorem~\ref{finiteAutomataStrong-effective}. All the following
theorems were announced first in~\cite{conf06}.

We say $f_n\to f$ for $f_n,f\colon\N\to\N$ if $\forall i\ \exists n\
\forall m > n\ f_m(i) = f(i)$.

\begin{theorem}\label{EAPprefixNumber}
Given $\omega\in\EAP$ and its regulator $f$, it is impossible to
compute algorithmically some $l\geqslant \p(\omega)$.
\end{theorem}

Remind that $\omega_T$ is the Thue--Morse sequence. This sequence
can be obtained as follows: let $a_0 = 0$, $a_{n + 1} = a_n\overline
a_n$, and $\omega_T = \lim_{n\to\infty}a_n$. Notice that $|a_n| =
2^n$. The Thue--Morse sequence has lots of interesting properties
(see~\cite{ThueMorse}), but we are interested in the following one:
$\omega_T$ is cube-free, i.~e., for any $a\in\B^*$, $a\ne\Lambda$
the string $aaa$ does not occur in $\omega_T$
(see~\cite{ThueMorse,Thue}).

\begin{proof}[Proof of Theorem~\ref{EAPprefixNumber}]
It is enough to construct $\omega_n\in\EAP$, $\omega\in\SAP$ with
regulators $f_n$, $f$ such that $\omega_n\to\omega$, $f_n\to f$, but
$\p(\omega_n)\to\infty$. Indeed, suppose the mentioned algorithm
exists and it outputs some $l\geqslant0$ (arbitrary for
$\omega\in\SAP$) given $\langle\omega, f\rangle$ on the input.
During the computation of $l$ the algorithm reads only finite number
of symbols in $\omega$ and of values of $f$. Hence there exists $N >
l$ such that algorithm does not know any $\omega(k)$ or $f(k)$ for
$k > N$. Since $\p(\omega_n)\to\infty$, there exists $n$ such that
$\p(\omega_n) > N$. The algorithm works on the input
$\langle\omega_n, f_n\rangle$ in the same way as it works on the
input $\langle\omega, f\rangle$, and then outputs $l$, but
$\p(\omega_n) > N > l$.

Let $\omega = \omega_T$, $\omega_n = a_na_na_n\omega$. Notice that
$\p(\omega_n)\geqslant2^n$. Indeed, if $\p(\omega_n) < 2^n$, then
$a_na_n\omega = a_na_na_n\overline a_n\overline
a_na_n\ldots\in\SAP$, and hence $a_na_na_n$ occurs in $\omega_T$~---
contradiction with the statement before the proof.

It only remains to show that we can find regulators $f_n$, $f$ for
$\omega_n$, $\omega$ such that $f_n\to f$. It is sufficient to find
the same regulator $g$ for all $\omega_n$ (then we can increase it
and obtain the same regulator for all $\omega_n$ and for $\omega$
too). Fix some $\r_\omega$ and assume $g = 4\cdot\r_\omega$. Let
$v$, $|v| = k$ occur in $\omega_n = a_na_na_n\omega$ infinitely many
times. Let us take the factor $\omega[i,j]$ of length
$4\cdot\r_\omega(k)$ and show that $v$ occurs in it. If $j\geqslant
3\cdot2^n + \r_\omega(k)$, then $v$ occurs on the factor
$\omega[3\cdot2^n, 3\cdot2^n + \r_\omega(k)]$ (by definition of
$\r_\omega$). Otherwise $j < 3\cdot2^n + \r_\omega(k)$, hence $i
\leqslant 3\cdot2^n - 3\r_\omega(k)$. But $i\geqslant0$, therefore
$\r_\omega(k) \leqslant 2^n = |a_n|$. Then $\omega_n[i, i +
\r_\omega(k)]$ is contained in $a_na_n$. But $a_na_n$ occurs in
$\omega$, so $\omega_n[i, i + \r_\omega(k)]$ occurs too. Therefore
$v$ occurs in $\omega$.

However $g$ is not already required. We should watch on the strings
occurring in $\omega_n$ finitely many times. Obviously, if some $v$
occurs in $\omega_n$ finitely many times, then $|v| = k > 2^n$
(otherwise $v$ occurs in two consequent strings $a_n$ or $\overline
a_n$, and thus in $\omega$). Therefore this can happen only for
finite number of different $n$. Considering all the situations when
strings of length $k$ occur in some $\omega_n$ finitely many times,
we probably increase the value $g(k)$, but only finitely many times.
Thus the required estimate for regulators exists.
\end{proof}

We have already seen that $\EAP\subsetneq\AP$
(Theorem~\ref{alPer-NotEssAlPer}). Using the same construction we
can show that it is impossible to separate these
classes~effectively.

\begin{theorem} \label{APseparateEAP}
Given $\omega\in\AP$ and its regulator $f$, it is impossible to
determine algorithmically whether $\omega\in\EAP$.
\end{theorem}

In~\cite{MuchSemUsh} the following universal method for construction
of strongly almost periodic sequences was presented. This method is
based on block algebra on words introduced in~\cite{Keane} and then
studied in~\cite{Jac}.

The sequence $\langle A_n, l_n\rangle$, where $A_n\subset\Sigma^*$
for some finite alphabet $\Sigma$, $l_n\in\N$, is called
\emph{strong $\Sigma$-scheme}, if the following conditions hold:\\[1mm]
(1) all the strings in $A_n$ have the length $l_n$;\\
(2) any string $u\in A_{n+1}$ has the form $u = v_1v_2\dots v_k$,
where $v_i\in A_n$, and for every $w\in A_n$ there exists $i$ such
that $v_i = w$.

\smallskip

We say that $\alpha\in\Sigma^\N$ is generated by strong
$\Sigma$-scheme $\langle A_n, l_n\rangle$ if for every $i$ and $n$
we have
  $$
 \alpha[il_n, (i+1)l_n - 1]\in A_n.
  $$

It is easy to see (by compactness) that any strong scheme generates
some sequence. In~\cite{MuchSemUsh}, the authors prove that any
sequence generated by strong scheme is strongly almost periodic.
Moreover every strongly almost periodic sequence is generated by
some strong scheme.

\begin{proof}[Proof of Theorem~\ref{APseparateEAP}]
It is enough to construct $\omega_n\in\EAP$,
$\omega\in\AP\setminus\EAP$ with the same regulator $f$ for all
$\omega_n$ such that $\omega_n\to\omega$.

In the same way as in Theorem~\ref{alPer-NotEssAlPer}, assume $a_0 =
1$, and then by the rule: $a_{n+1} = a_n\overline a_n\overline
a_na_na_n$. Denote $a_na_na_na_n$ by $c_n$. Put $l_n = 5^n-1 =
|á_0c_1\dots á_{n-1}|$. Consider $\omega = á_0á_1á_2á_3\dots$ and
$\nu = \lim_{n\to\infty} a_n$. From the proof of
Theorem~\ref{alPer-NotEssAlPer} it follows that
$\omega\in\AP\setminus\EAP$. Let $\omega_n = c_0c_1\ldots c_n\nu$.
The sequence $\nu$ is generated by the strong $\B$-scheme
$\langle\{a_n, \overline a_n\}, 5^n\rangle$, hence $\nu\in\SAP$.
Therefore $\omega_n\in\EAP$. Obviously, $\omega_n \to \omega$, and
it only remains to find common regulator $f$. We will get finite
number of conditions of the form $f(k) \geqslant \alpha$, then we
can take the maximum among all these~$\alpha$.

Let $v = \omega_n[i,j]$, $|v| = k$ occurs in $\omega_n =
c_0c_1\ldots c_n\nu$ infinitely many times. Then $v$ occurs in
$\nu$, hence in $a_m$ for some $m$ too. Therefore $v$ occurs in
$\omega$ infinitely many times and it is sufficient to take $f(k)
\geqslant \r_\omega(k) + \r_\nu(k)$.

Let $v = \omega_n[i,j]$, $|v| = k$ occurs in $\omega_n$ finitely
many times. Then $i < l_n$. Suppose $j > l_n$. If $k \leqslant 5^n$,
then $v$ occurs in $a_m$ for some $m$ and hence occurs in $\nu$
infinitely many times. But the inequality $k > 5^n$ holds only for
finite number of different $n$, and this yields just finitely many
conditions on $f(k)$. Now suppose $j \leqslant l_n$. But then $v$
occurs in $c_0c_1\ldots c_n$ and occurs in $\omega$ finitely many
times (otherwise $v$ occurs in $a_m$ for some $m$). Therefore in
this case it is sufficient to take $f(k) \geqslant \r_\omega(k)$.
\end{proof}

The following theorem shows that it is even impossible to separate
effectively~$\SAP$ and~$\Per$.

\begin{theorem} \label{SAPseparatePer}
Given $\omega\in\SAP$ and its regulator $f$, it is impossible to
determine algorithmically whether $\omega\in\Per$.
\end{theorem}

\begin{proof}
It is enough to construct $\omega_n\in\Per$,
$\omega\in\SAP\setminus\Per$ with common regulator $f$ for all
$\omega_n$ such that $\omega_n\to\omega$.

Every strongly almost periodic sequence can be obtained from the
strong $\Sigma$-scheme $\langle A_n, l_n\rangle$. Let us strengthen
the main condition on $A_n$: let us consider strong schemes such
that for each $n\in\N$ every $u\in A_{n+1}$ has the form $u =
v_1v_2\dots v_k$, where $v_i\in A_n$, and for any $w_1, w_2\in A_n$
there exists $i < k$ such that~$v_iv_{i+1} = w_1w_2$. Notice that
such schemes exist and can generate non-periodic sequences, e.~g.,
$\langle \{a_n, \overline a_n\}, 2^n\rangle$ generates~$\omega_T$.

Let $\langle A_n, l_n\rangle$ be the strong scheme satisfying the
strengthened condition from the previous paragraph, generating
$\omega\notin\Per$. Let $p_n = \omega[0,l_n]$. Thus $p_n\in A_n$ and
$\lim_{n\to\infty}p_n = \omega$. Assume $\omega_n =
p_np_np_n\ldots\in\Per$. Obviously $\omega_n \to \omega$ and it only
remains to find some common regulator $f$ for all~$\omega_n$.

Let $v = \omega_n[i,j]$, $|v| = k$ (since $\omega_n\in\Per$, it
follows that $v$ occurs in $\omega_n$ infinitely many times). The
inequality $k \geqslant |p_n| = l_n$ holds only for finite number of
different $n$, and this yields just finitely many conditions on
$f(k)$. Now we can assume that $k < l_n$. Take $t$ such that
$l_{t-1} < k \leqslant l_t$ (it is important that $t$ does not
depends on $n$ and is uniquely defined by~$k$). Then $t < n$. There
exists $m$ such that $ml_t \leqslant i$ and $j\leqslant (m+2)l_t$,
i.~e., $v$ occurs in some $ab$, where $a,b\in A_t$. Then by the
scheme property $v$ occurs in any $c\in A_{t+1}$. But on every
factor of $\omega_n$ of length $2l_{t+1}$ there exists an occurrence
of some $c\in A_{t+1}$ (fully contained in some $p_n$). Therefore it
is sufficient to take~$f(k) \geqslant 2l_{t+1}$.
\end{proof}

By the argument of Theorem~\ref{SAPseparatePer} we obtain that there
exists infinite set of periodic sequences with common regulator
(while the period tends to infinity). This construction can be used
in the following theorem: adding one symbol to the strongly almost
periodic sequence we can not check whether it is still strongly
almost periodic.

\begin{theorem} \label{find_pr_by_EAP}
Given $\omega\in\EAP$, its regulator $f$ and some
$l\geqslant\p(\omega)$, it is impossible to find algorithmically
$\p(\omega)$.
\end{theorem}

\begin{lemma} \label{whenSAPinPer}
If $a\omega\in\SAP$ for $a\in\Sigma^*$ and $\omega\in\Per$ with
period $l$, then $a\omega\in\Per$ with period $l$.
\end{lemma}
\begin{proof}
It is enough to prove lemma for single-letter $a$. Let $\alpha =
012\ldots(l-1)012\ldots(l-1)012\ldots(l-1)\ldots$ be periodic
sequence over alphabet $\Sigma_l = \{0, 1, 2,\dots, l-1\}$. Then by
Corollary~\ref{SAPtimesPer} we have $a\omega\times\alpha\in\SAP$. In
this sequence the symbol $\langle a, 0\rangle$ occurs infinitely
many times, hence $a = \omega(l)$.
\end{proof}

\begin{proof}[Proof of Theorem~\ref{find_pr_by_EAP}]
It is enough to construct $\omega_n\in\EAP$, $\omega\in\SAP$ with
common regulator $f$ for all $\omega_n$ such that
$\omega_n\to\omega$ and $\p(\omega_n) = 1$ ($\omega\in\SAP$ means
$\p(\omega) = 0$).

Notice that $1\omega_T\in\SAP$. Indeed, for each $n$ strings
$a_na_n$ and $\overline a_na_n$ occur in $\omega_T$, and hence
$1a_n$ too. Analogously $0\omega_T\in\SAP$.

By proof of Theorem~\ref{SAPseparatePer}, we can choose sequence
$k_n\to\infty$ such that all periodic sequences like
$\omega(0)\ldots\omega(k_n)\omega(0)\ldots\omega(k_n)\omega(0)\ldots$
have common regulator~$f$. Take a subsequence $m_n$ of the sequence
$k_n$ such that all the symbols $\omega(m_n)$ are equal. Suppose
these symbols are~0.

Let $\omega_n =
1\omega(0)\ldots\omega(m_n)\omega(0)\ldots\omega(m_n)\omega(0)\ldots$
and $\omega = 1\omega_T$. There exists common estimate $g$ on the
regulator for these sequences. Indeed, it is sufficient $g(k)
\geqslant f(k) + 1$ (by considering strings occurring infinitely
many times) and $g(k) \geqslant k$ (by considering strings occurring
only finitely many times: this can happen only for prefixes
occurring exactly once).

If $\omega_n\in\SAP$, then by Lemma~\ref{whenSAPinPer} we have
$\omega_n\in\Per$ with period $m_n$. But $\omega_n(0) = 1 \ne
\omega_n(m_n) = 0$. Therefore $\p(\omega_n) = 1$.

The case when all the symbols $\omega(m_n)$ are 1 is analogous (then
$\omega_n$ begins with~0).
\end{proof}

\section{Acknowledgements}

The author is grateful to A.~Semenov and An.~Muchnik for their help
in the work and also to M.~Raskin, A.~Rumyantsev, A.~Shen and to all
other participants of Kolmogorov seminar (Moscow) for useful
discussions. The results were also presented on the seminar under
the direction of S.~Adian, the author is thankful to all the
participants for attention.

\end{document}